# Nonlinear RF Fingerprints Authentication for OFDM Wireless Devices based on Demodulated Symbols

Yan Yan, Hong-lin Yuan *, Zhi-hua Bao and Guo-an Zhang

School of Information Science and Technology, Nantong University, Jiangsu Nantong 226019, China

(* Corresponding author, Email: yuan.hl@ntu.edu.cn, ORCID: 0000-0002-3320-9412)

**Abstract.** Radio Frequency fingerprints (RFF) authentication is one of the methods for the physical-layer information security, which uses the hardware characteristics of the transmitter to identify its real identity. In order to improve the performance of RFF based on preamble with fixed duration, a nonlinear RFF authentication method based on payload symbols is proposed for the wireless OFDM devices with the bit mapping scheme of QPSK. The communication system is modeled as a Hammerstein system containing the nonlinear transmitter and wireless multipath fading channel. A parameter separation technique based on orthogonal polynomial is presented for the estimation of the parameters of the Hammerstein system. The Hammerstein system parameter separation technique is firstly used to estimate the linear parameter with the training signal, which is used to compensate the adverse effect of the linear channel for the demodulation of the successive payload symbols. The demodulated payload symbols are further used to estimate the nonlinear coefficients of the transmitter with the Hammerstein system parameter separation technique again, which is used as the novel RFF for the authentication of the QPSK-OFDM devices. Numerical experiments demonstrate that the proposed method is feasible. The novel method can also be extended to the OFDM signals with other bit mapping schemes.

## 1 Introduction

As the rapid development of internet-of-things (IoT) and fifth generation mobile communication, the physical-layer security of wireless networks has inevitably become a hot topic. Radio Frequency fingerprints (RFF) authentication is one of the methods for the physical-layer information security, which relies on the hardware characteristics of the transmitter rather than the digital information such as cryptographic key or media access control (MAC) address to authenticate the real identity of the radio devices [1-3].

Although the priori digital preamble symbol of a communication frame is definite, the corresponding analog signals transmitted from different wireless devices are different as each transmitter hardware are unique, even though the devices are from the same type and same series. So, the received preamble signal is widely used to develop the RFF for the authentication of wireless devices. Ureten et al. [4] is the first published paper which uses the preamble signal to enhance the security of IEEE 802.11b wireless networks. In [5], the classification performance approaches the accurate rate of 80% when the signal to noise ratio (SNR) is greater than 6dB using three IEEE 802.11a devices with the power spectral density RFF of the preamble signals. The wavelet denoising and transform technique are used to extract the RFF from the preamble responses of IEEE 802.11a OFDM signals in [6]. In [7], the identification accurate rates of 28 different Wi-Fi devices are higher than 95% with the preamble RFF. Ramsey et al. [8] proposes a technique of preamble manipulation which can be used for the authentication of IEEE 802.15.4 devices accurately. However, the technique cannot differentiate devices with the same type. Liang et al. [9] proposes a RFF extraction technique based on empirical mode decomposition, which is effective to WLAN preamble signals when SNR is greater than 10dB. Jr et al. [10] proposes a technique for the wired device discrimination with the preamble signals etc., which provides a viable security augmentation for mitigating the vulnerability of networks. However, the preamble based RFF normally has short duration and is visible, it easily suffers the attacks of counterfeit with the high end instruments.

On the other hand, the nonlinearity of the transmitters is estimated as another important RFF, called nonlinear RFF in this paper, to authenticate the wireless devices. Nonlinear RFF of a radio device is determined by the power amplifier and digital to analog converter etc. of the transmitter, which includes the coefficient estimation of the nonlinear model of the transmitter [11-13] and mode decomposition or entropy transform [14-16] of the received signal etc. The nonlinear coefficients related RFF is normally estimated with the priori preamble and the corresponding received analog signal. In [11], the amplifier is modeled as a Taylor series, a multi-channel correlation RFF is proposed based on the derived carrier component and harmonic component expressions of the transmitted signal, which is independent to the base-band signal to be transmitted. The experiments of the classification with 4 FM emitters are successful. In [12], the wireless transmitter is modeled in a power-series expansion format which describes a memoryless nonlinear system. With the preamble as the input and the corresponding received signal passed through the wireless multipath channel and additive noise as the output, the coefficients of the nonlinear model and the multipath channel are estimated with an iterative algorithm. The estimated coefficients are considered as the nonlinear RFF to authenticate the radio device. In [13], a wide-band transmitter is modeled as a Hammerstein-Wiener system, and an improved niche genetic algorithm is proposed for the identification of the model. The estimated coefficients of the model are used as the nonlinear RFF to authenticate the transmitter. The simulation experiments show the good estimation precision and high classification rates. However, the duration of the preamble signal of a frame is normally limited in practice, which restrains the estimation performance of the relevant nonlinear RFFs.

In this paper, we estimate the nonlinear RFF with the demodulated payload symbols rather than the preamble symbols, which breaks through the limitation of the duration of the preamble symbols. The proposed nonlinear RFF authentication method is used for wireless OFDM communication with QPSK mapping. The channel includes additive white Gaussian noise (AWGN) and multipath fading channel together. The communication system is modeled as a Hammerstein system, and a Hammerstein system parameter separation technique is presented for the estimation of the parameters of the Hammerstein system based on [17-20].

## 2    System Model

The system model of the proposed nonlinear RFF authentication method for QPSK-OFDM wireless devices is shown in Fig. 1.

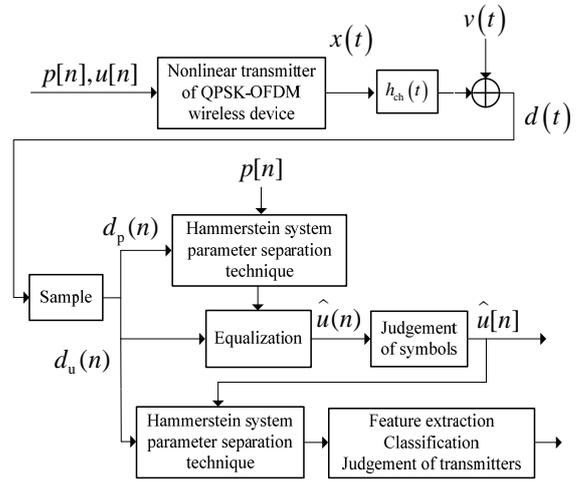

**Fig. 1.** The system model of the payload symbol-based nonlinear RFF authentication method for QPSK-OFDM wireless devices

In Fig. 1, $p[n]$ and $u[n]$ denote the frequency-domain (FD) training pilot and subsequent payload symbols of an OFDM frame, respectively. $x(t)$ is the emitted signal from the nonlinear transmitter of the wireless QPSK-OFDM device. $h_{ch}(t)$ denotes the impulse response of the wireless multipath fading channel. $v(t)$ is the additive white Gaussian noise (AWGN). The received signal $d(t)$ is sampled as discrete signal $d_p(n)$ and $d_u(n)$ successively, where $d_p(n)$ corresponds to the training pilot symbols and $d_u(n)$ corresponds to the payload symbols respectively. With $p[n]$ and $d_p(n)$, the finite impulse response (FIR) of the linear channel is estimated using the Hammerstein system parameter separation technique. And the estimated FIR is then used to equalize the payload signal $d_u(n)$. The equalized payload signal is denoted as $\hat{u}(n)$, which is judged as the received FD payload symbols $\hat{u}[n]$. $\hat{u}[n]$ and $d_u(n)$ are then used to estimate the parameters of the nonlinear channel with the Hammerstein system parameter separation technique again, which is used as the nonlinear RFF in this paper. The features are extracted from the obtained nonlinear RFF, and the classification and judgment of the transmitter entity are done with the extracted features.

## 3    The Hammerstein System Parameter Separation Technique

The nonlinearity of the OFDM transmitter is with memories. The QPSK-OFDM wireless transmitter and the channels are modeled as a Hammerstein system which contains a static nonlinear and a successive dynamic linear sub-system. The discrete equivalent model of the Hammerstein system is depicted in Fig. 2.

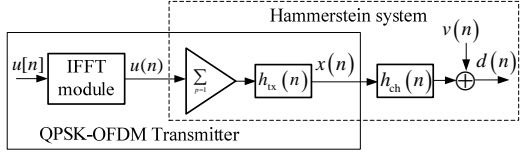

**Fig. 2.** The discrete equivalent model of QPSK-OFDM transmitter and channels

In Fig. 2, $u[n]$ is the FD QPSK symbols with the duration of $T_s$; IFFT module contains the operation of series-to-parallel, inverse discrete Fourier transform, adding cyclic prefix (CP) and parallel-to-serial; $u(n)$ denotes the discrete QPSK-OFDM preamble signal with the interval of $T_s$; $\sum_{p=1}$ denotes the static nonlinearity of the transmitter, and $h_{tx}(n)$ is the FIR corresponding to the memory property of the transmitter; $h_{ch}(n)$ denotes the FIR of the wireless multipath fading channel; $x(n)$ is the emitted signal of the transmitter; $v(n)$ is the AWGN, and $d(n)$ denotes the received wireless signal.

### 3.1 Linear vector model of the Hammerstein system

After the QPSK-OFDM signal $u(n)$ passes the static nonlinearity, it becomes

$$x_0(n) = \sum_{p=1}^{(P+1)/2} b_{2p-1}\phi_p(u(n)) \qquad (1)$$

where $P$ is an odd integer, $\phi_p(\bullet) = \bullet|\bullet|^{2(p-1)}$ is the conventional polynomial basis function, $b_i, i=1,3,\cdots,P$ are the corresponding coefficients. The emitted signal of the transmitter $x(n) = x_0(n) * h_{tx}(n)$ where * denotes the operation of linear convolution. The linear convolution of $h_{tx}(n)$ and $h_{ch}(n)$ is expressed as $h(n) = \sum_{l=0}^{L} h_l \cdot \delta(n-l)$, where $\delta(\bullet)$ is the unit impulse function and $L$ is the order of the cascade FIRs. Then, the received signal

$$\begin{aligned} d(n) &= x_0(n) * h(n) + v(n) \\ &= \sum_{l=0}^{L} \phi_{\mathbf{p},u}^T(n-l) h_l \mathbf{b} + v(n) \end{aligned} \qquad (2)$$

where $\phi_{\mathbf{p},u}(\bullet) = [\phi_1(u(\bullet)),\ldots,\phi_{\frac{P+1}{2}}(u(\bullet))]^T$ is the vector composed of the conventional polynomial basis functions of the QPSK-OFDM signal $u(\bullet)$, $\mathbf{b} = [1,b_3,\cdots,b_P]^T$ is the vector composed of the coefficients of the nonlinear conventional polynomial model. It can be seen with Eq. (2) that $d(n)$ without AWGN is the linear convolution of vector $\phi_{\mathbf{p},u}(n)$ and $h_n\mathbf{b}$. Suppose $z^{-1}$ denotes the unit delay, the linear vector model for the Hammerstein system is depicted as Fig. 3.

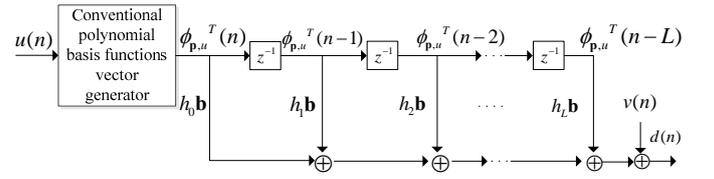

**Fig. 3.** The linear vector model for the Hammerstein system

Similar to [20], define the vector of QPSK-OFDM signal $\mathbf{u}_t = [u(1),\ldots,u(N)]^T$, $N$ is the number of $u(n)$; define the $l$-delayed vector of $\mathbf{u}_t$ is $\mathbf{u}_l = [\mathbf{0}_{1\times l}, u(1),\ldots,u(N-l)]^T$, $0 \le l \le L$, we have $\mathbf{u}_0 = \mathbf{u}_t$; define the conventional polynomial basis function of $\mathbf{u}_l$ is $\phi_p(\mathbf{u}_l) = [\mathbf{0}_{1\times l}, \phi_p(u(1)),\cdots,\phi_p(u(N-l))]^T$, $1 \le p \le \frac{P+1}{2}$; Then, define matrix $\mathbf{\Phi}_l = [\phi_1(\mathbf{u}_l),\cdots,\phi_{\frac{P+1}{2}}(\mathbf{u}_l)]$. Define the vector of the received signal $\mathbf{d} = [d(1),\ldots,d(N)]^T$, the vector of AWGN $\mathbf{v} = [v(1),\ldots,v(N)]^T$, the vector of the actual linear channel $\mathbf{h} = [h_0,h_1,\ldots,h_L]^T$, and the vector of the virtual channel combined of the actual nonlinear and liner channels $\mathbf{h_b} = [h_0\mathbf{b}^T, h_1\mathbf{b}^T,\ldots,h_L\mathbf{b}^T]^T$. Then

$$\begin{aligned} \mathbf{d} &= [\mathbf{\Phi}_0,\cdots,\mathbf{\Phi}_L][h_0\mathbf{b}^T,\ldots,h_L\mathbf{b}^T]^T + \mathbf{v} \\ &= \mathbf{\Phi}\mathbf{h_b} + \mathbf{v} \end{aligned} \qquad (3)$$

where $\mathbf{\Phi} = [\mathbf{\Phi}_0,\mathbf{\Phi}_1,\cdots,\mathbf{\Phi}_L]$. Then, the least square (LS)

estimation of $\mathbf{h}_b$ is $\hat{\mathbf{h}}_b = (\mathbf{\Phi}^H\mathbf{\Phi})^{-1}\mathbf{\Phi}^H\mathbf{d}$.

### 3.2 Hammerstein system parameter separation based on orthogonal polynomial

As the inversion of matrix $\mathbf{\Phi}^H\mathbf{\Phi}$ will often incur numerical instability caused by conventional polynomial, we use the orthogonal polynomial basis function in [19] to replace $\phi_p(\bullet)$ in $\phi_{\mathbf{p},u}(\bullet)$. Eq. (3) is then converted as

$$\begin{aligned}\mathbf{d} &= \mathbf{\Psi}\mathbf{h}_b^o + \mathbf{v} \\ &= \mathbf{\Psi}_{\mathbf{b}^o}\mathbf{h} + \mathbf{v}\end{aligned} \quad (4)$$

where $\mathbf{\Psi}_{\mathbf{b}^o} = \mathbf{\Psi}(\mathbf{I}_{(L+1)\times(L+1)} \otimes \mathbf{b}^o)$, $\mathbf{b}^o = \mathbf{U}^{-1}\mathbf{b}$, and $\mathbf{U}$ is an upper triangular matrix in [19] for orthogonal basis. With Eq. (4), the LS estimation of $\mathbf{h}$ is $\hat{\mathbf{h}} = (\mathbf{\Psi}_{\mathbf{b}^o}^H\mathbf{\Psi}_{\mathbf{b}^o})^{-1}\mathbf{\Psi}_{\mathbf{b}^o}^H\mathbf{d}$. As $b_1$ is supposed as 1 and $\hat{\mathbf{h}}_b^o = (\mathbf{\Psi}^H\mathbf{\Psi})^{-1}\mathbf{\Psi}^H\mathbf{d}$, $\hat{h}_0\mathbf{b}^o = \hat{\mathbf{h}}_b^o(1:\frac{P+1}{2})$ is obtained, then $\hat{\mathbf{b}} = \frac{\mathbf{U}\cdot\hat{h}_0\mathbf{b}^o}{\hat{h}_0\mathbf{b}^o(1)}$.

$\hat{\mathbf{h}}$ and $\hat{\mathbf{b}}$ are the separated linear and nonlinear parameters of the Hammerstein system.

## 4 The Proposed Nonlinear RF Fingerprints Authentication Method

Suppose the pilot arrangement of the QPSK-OFDM system is block type, which means FD symbols at all sub-carriers of the first OFDM symbol of a frame is the pilot symbols and the subsequent OFDM symbols of the frame are the payload symbols.

Let $\mathbf{d}_p$ and $\mathbf{d}_u$ denote the received discrete preamble and payload signal vector removing the CP, respectively. Let $\mathbf{p}$ denotes the FD pilot symbol vector. Using the Hammerstein system parameter separation technique with $\mathbf{p}$ and $\mathbf{d}_p$, we can obtain the estimation of the FIR of the combined linear channel of the QPSK-OFDM system, denoted as $\hat{\mathbf{h}}_p$, and the estimation of the nonlinearity parameter vector of the transmitter, denoted as $\hat{\mathbf{b}}_p$. And $\hat{\mathbf{b}}_p$ here is based on the pilot symbol and the corresponding preamble signal which is used for the performance comparison to the proposed payload symbol-based nonlinear RFF.

Define the fast Fourier transform (FFT) matrix as $\mathbf{F}$. Then, let us denote the vector of the $N$-FFT of $\hat{\mathbf{h}}_p$ as

$$\begin{aligned}\hat{\mathbf{H}} &= \mathbf{F}\hat{\mathbf{h}}_p \\ &= [\hat{H}_0,\cdots,\hat{H}_{N-1}]^T\end{aligned}. \quad (5)$$

Denote the $N$-FFT of $\mathbf{d}_u$ as

$$\begin{aligned}\mathbf{D}_u &= \mathbf{F}\mathbf{d}_u \\ &= [D_0,\cdots,D_{N-1}]^T\end{aligned}. \quad (6)$$

As the CP removal transfers the linear convolution channel to the circular one, the FD one-tap equalized vector of the $N$-FFT of $\mathbf{d}_u$ is written as

$$\hat{\mathbf{u}}_e = [\frac{D_0}{\hat{H}_0},\cdots,\frac{D_{N-1}}{\hat{H}_{N-1}}]^T. \quad (7)$$

The judged vector of $\hat{\mathbf{u}}_e$ is denoted as $\hat{\mathbf{u}}_f$ which is the demodulated FD payload QPSK symbols. Using the Hammerstein system parameter separation technique with $\hat{\mathbf{u}}_f$ and $\mathbf{d}_u$, we can obtain the estimation of the nonlinear parameter vector of the transmitter, denoted as $\hat{\mathbf{b}}_u$, which is the novel nonlinear RFF of the wireless device based on the demodulated payload symbols.

## 5 Numerical Experiments

The numerical experiments with two nonlinear transmitters and a multipath fading channel are done to verify the feasibility of the proposed method.

The number of sub-carriers of an OFDM symbol is set to 2048 and the bit mapping scheme for FD symbols on all sub-carriers is QPSK. The length of CP is set to 512, and no virtual sub-carriers are employed. The number of the training OFDM symbol of a frame is 1, and the number of the payload OFDM symbols in the frame is set to $p$ = 1, 2, 4, and 8, respectively. The static nonlinear parameters of the two transmitters [12] to be authenticated are shown in Table 1.

**Table 1.** Static nonlinear parameters of two transmitters

| Coefficient | Transmitter-1 | Transmitter-2 |
| --- | --- | --- |
| $b_1$ | 1 | 1 |
| $b_3$ | $-0.0735 - i*0.0114$ | $-0.0910 + i*0.1580$ |
| $b_5$ | $-0.0986 + i*0.0590$ | $0.2503 + i*0.0286$ |
| $b_7$ | $-0.0547 - i*0.0055$ | $0.0155 + i*0.0025$ |

Unlike a fixed 3-path empirical channel used in the numerical experiments of [12], we apply a random Rayleigh fading channel with maximum delay of 8 samples and path number of 5 to emulate the combined linear channel containing the memory effect of the transmitter and the actual wireless multipath channel. The channel is supposed as constant over one transmission.

Then, the MontCarlo classification experiments for the two transmitters are done with $\hat{\mathbf{b}}_u$ and $\hat{\mathbf{b}}_p$ respectively when $E_b/N_0$ is set from 0dB to 20dB with 5dB as the step. The second coefficient of $\hat{\mathbf{b}}$, denoted as $\hat{b}_3$, is extracted as the feature of the transmitter to be classified. The times of the MontCarlo experiments are set to 100, and 66 samples of $\hat{\mathbf{b}}$ are observed for each transmitter in each experiment.

When $E_b/N_0$ is set to 10dB and the number of the payload and preamble OFDM symbol are all set to one, one group of samples of $\hat{b}_3$ from $\hat{\mathbf{b}}_u$ and $\hat{\mathbf{b}}_p$ for two transmitters are shown in Fig. 4, where $\text{Re}(\hat{b}_3)$ and $\text{Im}(\hat{b}_3)$ denote the real and imagery part of $\hat{b}_3$ respectively.

It can be seen from Fig. 4 that the feature samples from $\hat{\mathbf{b}}_u$ and $\hat{\mathbf{b}}_p$ can be classified to certain extent. It also can be seen that the separability of the feature samples from $\hat{\mathbf{b}}_u$ is worse than that from $\hat{\mathbf{b}}_p$ with the same $E_b/N_0$ and OFDM symbol number, which is caused by the error bits during the demodulation of the payload symbols.

However, when the number of the payload OFDM symbols increases, the separability of the feature samples from $\hat{\mathbf{b}}_u$ becomes better. One group of samples of $\hat{b}_3$ from $\hat{\mathbf{b}}_u$ for the two transmitters with eight demodulated payload OFDM symbols are shown in Fig. 5 when $E_b/N_0$ is also 10dB, and the samples from $\hat{\mathbf{b}}_p$ with one preamble symbol in Fig. 4 are shown for comparison with the same *x* and *y* axis.

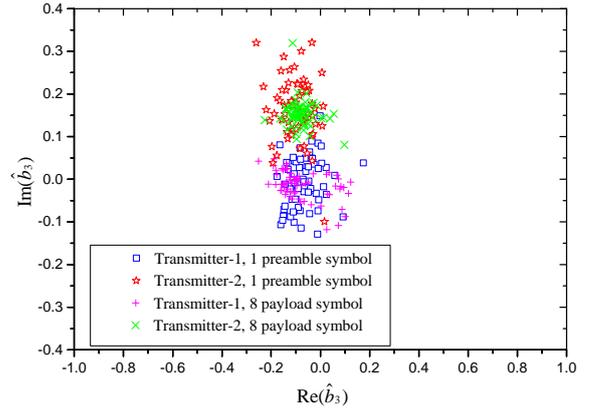

**Fig. 5.** One group of samples of the features from $\hat{\mathbf{b}}_u$ for two transmitters with 10dB $E_b/N_0$ and eight payload symbols

With Fig. 5, it can be seen that the feature samples from $\hat{\mathbf{b}}_u$ with eight demodulated OFDM payload symbols are clustered well, and their separability is better than that based on one preamble symbol.

A standard *k*-NN classifier is used to classify the two transmitters with the feature $\hat{b}_3$ from $\hat{\mathbf{b}}_u$ and $\hat{\mathbf{b}}_p$ respectively. For one time of classification experiment, 33 feature samples are used as the training set, and the other 33 samples are used as the testing set. The correct classification rates of 100 times MontCarlo experiments are averaged. The results with $\hat{b}_3$ from $\hat{\mathbf{b}}_u$ and $\hat{\mathbf{b}}_p$ are shown in Fig. 6

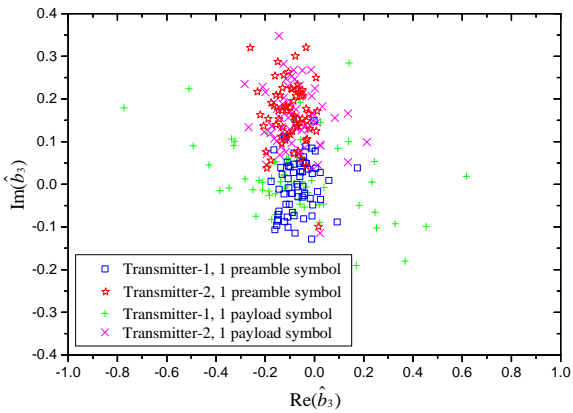

**Fig. 4.** Samples of the features from $\hat{\mathbf{b}}_u$ (payload symbol-based) and $\hat{\mathbf{b}}_p$ (preamble symbol-based) for two transmitters with 10dB $E_b/N_0$ and one symbol

where $k = 3$, the number of the preamble OFDM symbol is 1, and the number of the payload OFDM symbols is $p = 1, 2, 4,$ and 8, respectively.

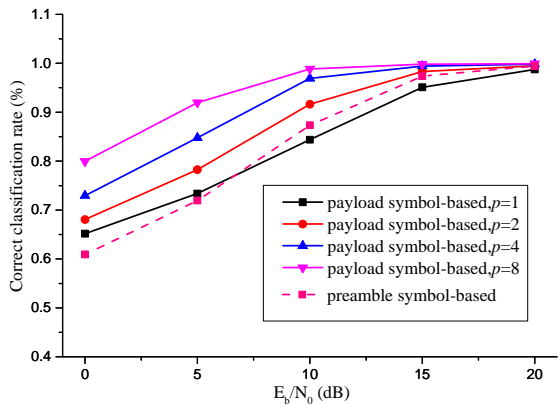

**Fig. 6.** The correct 3-NN classification rates in different $E_b/N_0$ for the preamble symbol-based and payload symbol-based nonlinear RFF

It can be seen from Fig. 6 that the correct classification rates increase with the increase of $E_b/N_0$. Fig. 6 also shows that the correct classification rates with payload symbol-based nonlinear RFF when $p$=2,4,8 are bigger than that with preamble symbol-based nonlinear RFF at all $E_b/N_0$. As the increase of the number of the payload OFDM symbols of a frame, the correct classification rates increase.

On the other hand, it can be seen from Fig. 6 that the classification performances based on one payload symbol are near to that based on the preamble symbol. And the correct classification rates with the two kinds of RFF are greater than the random guess probability of the two transmitters even when $E_b/N_0$ is 0dB. When the number of payload symbols is 8, the correct rates reach 80% when $E_b/N_0$ remains 0dB. When $E_b/N_0$ is greater than 10dB and the number of payload symbols is 8, the correct classification rates are close to 100%. And when $E_b/N_0$ is 20dB and the number of payload symbols is one, the correct classification rates are also near to 100%.

## 6 Conclusion

In this paper, we have proposed a nonlinear RFF authentication method for QPSK-OFDM wireless devices based on the demodulated payload symbols, rather than the known training symbols. A Hammerstein system parameter separation technique has also been presented, which is used twice for the proposed method. Numerical simulations show that the longer the length of the payload symbols, the better the correct classification performance of the proposed nonlinear RFF, which is not limited by the duration of the preamble signal any more. The novel method can also be extended to the OFDM signals with other bit mapping schemes.